# Surveying Turkish high school and university students' attitudes and approaches to physics problem solving

Nuri Balta[1], Andrew Mason[2] and Chandralekha Singh[3]
[1] *Canik Basarı University, Faculty of Education, Samsun, Turkey 55080*
[2] *University of Central Arkansas, Department of Physics, Conway, AR 72035*
[3] *University of Pittsburgh, Department of Physics and Astronomy, Pittsburgh, PA 15260*

**Abstract.** Students' attitudes and approaches to physics problem solving can impact how well they learn physics and how successful they are in solving physics problems. Prior research in the US using a validated Attitude and Approaches to Problem Solving (AAPS) survey suggests that there are major differences between students in introductory physics and astronomy courses and physics experts in terms of their attitudes and approaches to physics problem solving. Here we discuss the validation, administration and analysis of data for the Turkish version of the AAPS survey for high school and university students in Turkey. After the validation and administration of the Turkish version of the survey, the analysis of the data was conducted by grouping the data by grade level, school type, and gender. While there are no statistically significant differences between the averages of various groups on the survey, overall, the university students in Turkey were more expert-like than vocational high school students. On an item by item basis, there are statistically differences between the averages of the groups on many items. For example, on average, the university students demonstrated less expert-like attitudes about the role of equations and formulas in problem solving, in solving difficult problems, and in knowing when the solution is not correct, whereas they displayed more expert-like attitudes and approaches on items related to meta-cognition in physics problem solving. A principal component analysis on the data yields item clusters into which the student responses on various survey items can be grouped. A comparison of the responses of the Turkish and American university students enrolled in algebra-based introductory physics courses shows that on more than half of the items, the responses of these two groups were statistically significantly different with the US students on average responding to the items in more expert-like manner.

**PACS:** 01.40Fk, 01.40.gb, 01.40G

## I. INTRODUCTION

Prior research suggests that students in introductory physics courses have epistemological beliefs about physics and learning physics that are often very different from those of physics experts [1-6]. These studies point to the important fact that such differences in expert-novice attitudes towards physics and learning physics can impact what students actually learn in their physics courses. Other studies suggest that problem solving skills must be developed while learning physics, i.e., one must develop a good knowledge structure of physics principles and concepts while also simultaneously developing useful skills to solve problems in different contexts [7-13]. Indeed, in order to become a physics expert and learn to think like a physicist, there should be growth along all these intimately intertwined dimensions: gradual construction of a robust knowledge structure, development of problem solving skills, and development of attitudes towards physics learning and problem solving that are expert-like.

Much of the research literature pertaining to the US physics students' epistemological beliefs about physics and physics learning uses Likert-scale surveys, consisting of statements with which a respondent may agree or disagree. For example, the Maryland Physics Expectation Survey (MPEX) [4] and the Colorado Attitudes about Science Survey (CLASS) [5, 6] are well-known surveys that explore students' epistemological beliefs about physics and physics learning. They are usually administered at the beginning and also at the end of instruction in an introductory physics course. These surveys have identified certain persistent trends.

For example, students in introductory physics courses, who are still relative novices, often perceive physics knowledge to be a collection of disconnected facts and formulas to memorize, rather than having a coherent structure. Many introductory students appear to have not internalized the hierarchical organization of physics knowledge and may not see the need to focus on building such a hierarchical knowledge structure. Moreover, many introductory students begin a physics course believing that they are unable to truly learn physics since only those with sufficiently high intelligence can learn physics. [4,6]. Such unproductive epistemological beliefs can negatively impact learning.

According to the MPEX data, students' overall epistemological beliefs about physics and learning physics in traditionally taught introductory physics courses deteriorate somewhat by the end of instruction compared to at the beginning of instruction [4]. Furthermore, when introductory students were asked to answer CLASS survey questions from their own and their professors' perspectives, they seemed to be able to respond in a more expert-like manner when considering their professors' point of view, but maintained a less expert-like stance when answering from their own point of view [6]. These trends in student epistemology towards physics and learning physics are difficult to overcome, and relatively few, carefully designed introductory physics courses and curricula are able to produce a positive change in these attitudes [14–16].

While instruments such as the MPEX and CLASS address a broad scope of epistemological beliefs about physics and learning physics, other instruments have chosen to focus on attitudes towards specific aspects of learning physics. For example, Cummings et al. [14,17] developed the Attitudes toward Problem Solving Survey (APSS) from the MPEX survey in order to focus specifically on students' attitudes toward physics problem solving [17]. When administered in introductory physics classrooms in three different institution types, the APSS results showed that students' attitudes about problem solving did not improve after instruction, and that attitudes were least expert-like (least favorable) at the large university with a large class size.

Considerations of both attitudes and approaches towards physics problem solving can impact students' success in physics problem solving. Therefore, the Attitudes and Approaches to Problem Solving (AAPS) survey, a modified version of the APSS survey [18,19], was developed, validated and administered to physics faculty, graduate students and introductory students in physics and astronomy courses. After the initial development and validation, further validation of the AAPS survey including its content and face validity was conducted with a wide spectrum of university-level students, from introductory-level freshmen to physics graduate students and faculty at a research university in the United States [19].

The AAPS survey contains additional questions compared to APSS and showed that in traditionally taught classes, introductory physics and astronomy students generally have significantly less favorable attitudes and approaches to problem solving for introductory level problems compared to physics faculty and graduate students [18,19]. The survey also shows that on several dimensions, graduate students have significantly less favorable attitudes and approaches to problem solving compared to physics faculty for graduate-level physics problem solving. For these comparisons, "favorable" was defined based upon the responses of experts similar to the earlier surveys [4,6,7].

In addition to lack of sufficient prior experience with physics problem solving and the broad spectrum of preparatory background of students in the introductory physics courses, the significantly less favorable responses to the AAPS survey can also partly be attributed to the traditional teaching approach [1,4]. In a traditional instructional setting, the instructor is generally a "sage on the stage" authority, whose job is to impart the physics knowledge using a teaching by telling paradigm, and the students' responsibility in turn is to take notes, memorize content, and restate those things on exams and other assignments. This type of instructional approach is not conducive to learning physics and developing a robust knowledge hierarchy and superior problem solving skills, and is inimical towards positive attitudes and approaches. Traditionally taught students are more likely to assume that if they cannot solve a problem within 10 minutes they are not smart enough to learn physics and such an assumption may interfere with the likelihood that these students will make an effort to explore effective strategies for solving problems and learning physics [4].

Improved attitudes and approaches to problem solving [14-16] often result from strategies that engage students actively in the learning process, e.g., peer instruction [24,25], collaborative group problem solving [26,27] and modeling problem solving in a realistic way such that students understand how the professor must also struggle with a challenging problem if not familiar with the solution [28,29], etc.

Another important consideration is how the attitudes and approaches to problem solving for students in the US differ from those in other countries who are exposed to different social and cultural norms and learn physics in different types of educational systems with different types of instructional constraints and affordances. In this paper, we will discuss the validation, administration and analysis of data for a Turkish version of the AAPS survey for high school and university students in Turkey. The analysis of data was conducted by grade level, school type, and gender. The survey was administered to high school students from two different grade levels and three different types of high schools and also to university students in introductory algebra-based physics courses. We examined trends between different class levels, between different types of schools, and between male and female students on survey responses. The comparison of the responses of the Turkish students enrolled in an algebra-based university introductory physics course with responses of the American students enrolled in algebra-based physics courses in the US is also presented. An exploratory factor analysis is also conducted to investigate the natural dimensions along which student responses fall.

## II. METHOD

### A. Validation of the Turkish Translation of the AAPS Survey

An expert (professional) science translator translated the AAPS survey from English to Turkish. The translation was first evaluated by two science and two English faculty members who could speak and read both Turkish and English well. It was then administered to four high school students during a one-on-one administration to ensure that the wording was understood as intended, even by 10$^{th}$ and 11$^{th}$ grade students taking physics (more details are provided in the instrument subsection below). The Turkish version of the AAPS survey was also shown to five additional physicists who could speak and read both Turkish and English well. Comments from the experts and students were used to make small changes to the Turkish version of the AAPS survey and further validate the translated version.

### B. Participants for Large Scale in-class Turkish Study

This study was carried out with a total of 528 high school and university students in Turkey. The instruction at all levels was traditional (lecture and labs) according to the American standards. In Turkey, high school lasts for four years from grades 9-12 corresponding to students with ages between 15-18 years. Physics is a mandatory subject for 9$^{th}$ and 10$^{th}$ grade students and students intending to choose a career related to science take physics in 11$^{th}$ and 12$^{th}$ grades. All high school students taking physics use the same national curriculum and textbook mandated by the government. There are various types of high schools but three types of high schools dominate: regular (generally called Anatolia school), science and vocational schools. Since science is generally considered a passport to a successful career, most students (and their parents) want to pursue science. In particular, after the eighth grade level, all students take a high-stakes exam (abbreviated as TEOG). Students scoring highest on this exam generally enroll in science high schools, those who obtain moderate scores enroll in regular high schools and students who obtain low scores attend vocational high schools. The number of science high schools is very small compared to regular and vocational high schools. Furthermore, the number of students in science high schools is also limited and all students in these schools choose many of the science subjects such as mathematics, physics, chemistry and biology. In particular, these science high schools were established to provide education to the exceptionally gifted mathematics and science students. The goal of regular high schools is to provide students with a balanced education in liberal arts, as well as to prepare them for higher education and also for careers in which they may be interested. Vocational high schools prepare qualified students for various professions or for higher education in specialized areas. For example, vocational high schools offer courses such as communications technology, ceramics, electrical science and engineering-electronics, food technology, technical drawing, and library science, which may not be typically available to those enrolled in most of the science and regular high schools.

There are approximately 36 weeks in one academic year in Turkish high school. There are two weekly physics course hours in all school types in $9^{th}$ and $10^{th}$ grades. In $11^{th}$ and $12^{th}$ grades, there are four weekly physics course hours in regular and science high schools. (There are no physics courses in vocational schools in $11^{th}$ and $12^{th}$ grades; however, students may take courses such as electric and electronics engineering which involves physics. Therefore, students in vocational school took the AAPS survey assuming it was in the context of problem solving in such a course.). Moreover, the physics curriculum does not change across schools. In Turkey, all schools at all grade levels, in all subjects (physics, chemistry, etc.), have to use the books prepared or certificated by the Ministry of National Education (MoNE) for that grade in a particular subject. Although it is not compulsory, teachers are free to choose one or more resource books in addition to the book certified by MoNE. Moreover, although the curricula for all school types and subjects are prepared by the MoNE, private schools can use a different curriculum and use different books, provided that the MoNE certifies it as comparable to their textbook and curriculum. We note that among the six high schools that participated in this study, two of them were private high schools but they used the same curriculum and text books as public schools licensed by MoNE.

The university students who participated in this study were first year students majoring in science education at the university (there is no equivalent major in the US since these Turkish students pursue this major immediately after high school). Once these students graduate from university, they typically teach science in middle schools (the middle schools have separate science teachers in Turkey). Therefore, during their four years at the university, along with courses related to educational sciences, they must take two semesters each of introductory university physics, university chemistry and university biology. To enroll in the science education program at the university, these students must pass the two consecutive compulsory national high stakes exams (abbreviated as YGS and LYS) at the end of the $12^{th}$ grade level, in which they must solve high school level physics problems (along with math, chemistry, etc.). We note that in Turkey, once students finish four years of high school, they all take the YGS and LYS exams. If they succeed in these exams with very good scores, they generally continue their education at a university in the major of their choice, but if they score well only in the first exam (YGS), they generally enroll in a college. Colleges are two-year institutions, while universities are four-year institutions and most also have graduate programs. While students graduating from regular and science high school strive to continue their education mostly in Turkish universities, the vocational school graduates generally go to Turkish colleges if they do not take a job right after high school. Thus, the university students in this study majoring in science education are those who either graduated from a regular high school or a science high school. The material for the two-semester university algebra-based course is similar to the material for $11^{th}$ and $12^{th}$ grades in high school (in this sense, there is repetition of content in high school and university), but the university courses are generally more intensive since students have three course hours of physics each week for 28 weeks each year.

When the Turkish university students took the AAPS survey, they had already completed university algebra-based mechanics in the fall semester and were enrolled in the university algebra-based electricity and magnetism course in the spring semester together with the general chemistry and general biology courses both semesters. Since all science instructors for science education majors communicate and teach together, due to the convenience of instructors, the AAPS survey was administered to students during a general chemistry class at the end of the spring semester. Since it was their first year after high school, students were generally 19-20 years old. Students were not given any grades or bonus scores for answering the survey questions but were told that this is part of a research study to improve education. When the importance of a research survey to improve education is well explained, it has been observed in the past that there is generally honest student participation on the surveys in Turkey.

This study was carried out in two different cities in Turkey and data were collected from all six high schools (two regular, two science and two vocational) and from one university. In other words, instead of sampling students in many more high schools and universities in Turkey, a convenience sampling procedure was used, which involved students in schools easily accessible to researchers. This technique was preferred to achieve a reasonable sample size in a reasonable time in an inexpensive way since the researchers knew the high school teachers whose students participated in the study. Students in grade 12 did not participate in this research study since their high school teachers generally do not involve them in any research studies or other activities as their time and energies are directed exclusively towards preparing for national university entrance examinations (YGS and LYS). Therefore, only grade 10 and grade 11 students (typically aged 15-17 years) participated in this study.

The demographics of the 528 students who participated in this study are presented in Table I, along with the numbers of students in each group (frequencies) and associated percentages. The sum of the numbers of students in the groups generally does not add up to 528 because some students did not provide the full information to place them in a particular group. For example, in Table I, there are 346 male and 171 female students, which add up to 517 students, indicating that 11 students did not report their gender. A similar situation applies to other groups.

Table I. Number of Turkish high school and university students who were administered the AAPS survey by gender, grade level, and school type.

|  | N | % |  | N | % |
|---|---|---|---|---|---|
| **Gender** | | | **School** | | |
| Men | 346 | 66 | Regular school | 192 | 37 |
| Women | 171 | 32 | Science school | 102 | 19 |
| Not specified | 11 | 2 | Vocational school | 187 | 35 |
| **Grade** | | | University | 40 | 8 |
| 10th | 280 | 53 | Not specified | 7 | 1 |
| 11th | 186 | 35 | | | |
| University | 43 | 8 | | | |
| Not specified | 19 | 4 | | | |

As shown in Table I, the percentages of participants that omitted demographics related to gender, grade level, and school type for each group range from 1% to 4%. Therefore, the number of missing values in each group does not impact the representative class sample (as noted, we removed the students who did not provide data related to a particular demographic in the analysis focusing on that demographic category, e.g., if a student did not provide gender information, we did not include that student in the analysis based on gender) [23].

**C. Survey Instrument**

The Attitudes and Approaches to Problem Solving (AAPS) survey, which was developed, validated, and administered by Mason and Singh in the US to a wide-range of students and physics faculty [19], is an inventory of 33 Likert type items. The survey questions are organized in the form of statements that one could agree or disagree with on a scale of 1 (strongly agree) to 5 (strongly disagree) with 3 signifying a neutral response. The items are worded such that while for 24 of them, "strongly agree" and "agree" are favorable responses, for nine items "strongly disagree" and "disagree" are favorable responses (expert-like responses).

As noted earlier, since the original AAPS survey was developed and validated in English, we first translated and validated the Turkish version of the instrument with both experts and students before administrating it in high schools and university course in Turkey. The AAPS survey was translated into Turkish by an expert science translator with the help of two science education researchers, and the translation was examined by two English language teaching instructors at Canik Basari University. The translation process was initially conducted via email between the expert translator and one of the science education researchers until a consensus was established. Then, another researcher checked the translation and two English language instructors (Turkish native speakers) went over the version and provided feedback. The survey was tweaked slightly based upon their comments. The survey was shown to five other physicists with knowledge of both Turkish and English who either reside in US or were visiting a US university from Turkey, three of whom did not have any suggested changes and two had minor suggestions. The Turkish version of the survey at this point was one- to- one with the English version with one exception: on item 9, conservation of linear momentum was changed to Newton's second law since $10^{th}$ grade students had instruction in this topic but not in conservation of linear momentum.

At this point, the survey was administered to four high school students in a one-on-one situation to make sure that students interpreted the questions unambiguously. We got feedback from high school students because the researchers determined that high school students who have the least experience with physics learning are more likely to point to any wording that is unclear to them. First, two of them (one from $10^{th}$ and one from $11^{th}$ grade) were asked to respond to the questions including writing their views on what each item meant to them and if they had any difficulty comprehending any of the survey questions. One student noted that the seventh item, which is composed of a long sentence, was somewhat difficult (although still possible) to understand. The researchers checked both the English and Turkish versions of the item to ensure it was understandable, and decided to keep a slightly tweaked version of the question since it measures an important concept related to students' problem solving approaches (it identifies whether to be able to use an equation to solve a problem, the student thinks about what each term in the equation represents and how each term matches the problem situation). Moreover, the item-total correlation was reasonably good (0.413) for the seventh item. These students also provided valuable feedback that guided the researchers to replace the literal formal Turkish words for "approach", "conceptual" and "intuitive" on the survey questions with more understandable common synonymous words in Turkish. After tweaking the survey, we had two other students (one from $10^{th}$ and one from $11^{th}$ grade) respond to the AAPS survey questions while thinking aloud. We observed that they answered all items without hesitation and were able to interpret all questions correctly. For three items (different ones for different students), the students read the items two times but completed the survey easily without any difficulty. We asked students at the end why they read those questions twice, e.g., why one of them read the third item two times. The student replied that she did not grasp the meaning of "being able to handle the mathematics" in the first reading but had no problem with the wording when she read it again. After both one-on-one interviews with these students, minor changes were made. Since AAPS survey items were easily understandable to Turkish high school students, the researchers administered it to 528 students in various classes in various high schools and a university.

The AAPS survey was administered to all student groups at the end of the second semester of the 2014-2015 school year. The sheets comprising the AAPS survey items were delivered to the physics instructors (including the university instructor). The high school teachers administered the survey during their physics courses but the university instructor administered it in a general chemistry class since that time worked out to be the most convenient (the physics and chemistry university instructors for the science education majors had a history of working together and taught the same science education students). Instructors were told to allow their students the last 20 minutes of the class time for responding to the survey. The instructors explained to students that the survey was for research purposes to improve education, but they were not given any course bonus points for taking the survey. Past experiences suggest that in Turkey, students generally take the surveys honestly when they are told that it is for research purposes to improve education. We note that the instructors who administered the survey in their classes were requested to report any queries from students regarding the understandability of the items. This was done to ensure that the Turkish version which had been iterated with several experts and four high school students was understood unambiguously by all students. All of the instructors who administered the survey noted that none of their students asked them to clarify any issues on the survey.

Once the data were collected, to check the reliability of the Turkish version of the AAPS survey, the Cronbach α test for internal consistency was calculated on the raw Likert scale data [29,30]. Table II shows the values for α for all students for various groups. For the entire student cohort and also for each student group based upon different criteria (gender, type of school and grade level), the Cronbach alpha statistic is robust ($\alpha > 0.80$) and ranges from 0.82 to 0.90 across different student groups [30]. These values are comparable to the values reported by Mason and Singh for the English version of the survey administered to students from the introductory to graduate levels and provide further validity to the Turkish version of the survey.

Table II. Cronbach α statistics for different populations of students who were administered the AAPS survey along with the number of students.

| Category | All | Grade | | School | | | | Gender | |
|---|---|---|---|---|---|---|---|---|---|
| | | 10th | 11th | Regular | Science | Vocational | University | Male | Female |
| N | 528 | 280 | 186 | 192 | 102 | 187 | 40 | 346 | 171 |
| α | 0.852 | 0.817 | 0.897 | 0.842 | 0.838 | 0.876 | 0.831 | 0.859 | 0.845 |

### D. Data analysis

Initially, factorial ANOVA was conducted to investigate the overall effect of gender, school type and grade level on students' attitudes and approaches to problem solving [30-32]. However, one of the assumptions of the ANOVA was not met. The "Levene's Test of Equality of Error Variances" showed that the error variance of the dependent variable (students' scores) was not equal across groups. This means that conducting factorial ANOVA was not as reliable as a nonparametric chi-squared test [30-32]. Thus, a nonparametric test (chi-squared test) was conducted, based on frequency distributions of responses among groups. In particular, chi-square tests were used to assess whether the frequency distribution of responses to the AAPS survey items between different groups of participants (e.g., males versus females; 10th grade versus 11th grade students, and students grouped by school type) were statistically different or not. The chi-square test was also used to find out whether there were statistically significant differences between groups for each of the 33 individual survey questions and for comparing responses by American and Turkish students, who were enrolled in algebra-based introductory physics courses. A principal component analysis was also conducted to investigate the natural clusters in which student responses to the survey items can be grouped.

## III. RESULTS FROM LARGE SCALE TURKISH IMPLEMENTATION

The collected data were categorized in terms of gender, grade level and type of high school. No statistically significant differences were observed between the overall means of the groups on the entire AAPS survey responses taken together. Even though the overall differences between groups were not statistically significant, on an item by item basis, there are some interesting differences. Both descriptive and inferential statistics were used to assess the variances on item by item basis. For descriptive statistics for each item, the normalized data were used. To normalize the data, first, for each item, a "+1" is assigned to each favorable (expert-like) response, a "−1" is assigned to each unfavorable response, and a 0 is assigned to neutral responses (denoted by 3 on a 1-5 Likert scale) [19]. Each respondent has a score of -1, 0 or +1 for each item after this process. The group normalized average score for a particular survey item can be calculated after the number of favorable, unfavorable or neutral responses for all students in the group is tabulated. In particular, the normalized score for a particular survey item for a group can be obtained by summing over the number of positive (+1), negative (-1) and neutral (0) responses and dividing by the total number of responses [19]. We note that if the number of students with a +1 score on that item is larger than the number of -1 score, the response of the group on that item on average is more expert-like (and will be denoted by a positive number. Similarly, a normalized score with a negative sign (as in the example for male students on item 1) denotes an overall unfavorable response. The entire AAPS survey [19] and the favorable and unfavorable responses to each item can be found in EPAPS [36].

In the following sections, the normalized statistics for all items for Turkish students are initially presented across gender, grade level and type of school, and then separately analyzed for some particularly interesting items. For inferential statistics, chi-square analysis was conducted to investigate significant statistical differences between groups, including differences in average responses of the Turkish students in an algebra-based university physics course as compared with US students in algebra-based courses.

### A. Descriptive and inferential analysis on the basis of groups

The average numbers of favorable, unfavorable and neutral responses for each group on the entire survey were computed and converted to average percentage values as shown in Fig. 1. For every group shown in Fig. 1, the average percentages of favorable responses on the entire survey are more than that of unfavorable responses. In terms of gender, no statistically significant differences were found between male and female students on the average percentage of favorable responses (or unfavorable responses).

Fig. 1 also shows that for type of schools, the average percentage of expert-like responses is highest for university students (50%) and lowest for vocational high school students (42%). One-way ANOVA was conducted to investigate differences between the groups based upon type of schools in terms of favorable responses and the F test results were significant ($F(3, 517) = 3.548$, $p = 0.014$). However, further analysis (Bonferroni correction [31-32]) showed that the statistically significant difference was only between the average favorable responses of university and vocational high school students ($p=0.010$). This implies that in terms of the average favorable responses, while regular high school, science high school and university students do not differ statistically significantly, university students are significantly more expert-like than vocational high school students in their responses. This finding is reasonable considering vocational high school students were placed in that group because they had performed worst in their national standardized test (which included science) at the end of $8^{th}$ grade and there may be a correlation between attitudes and approaches to problem solving and their science performance. In terms of the average unfavorable responses, no statistically significant differences were observed ($p=0.261$) between groups.

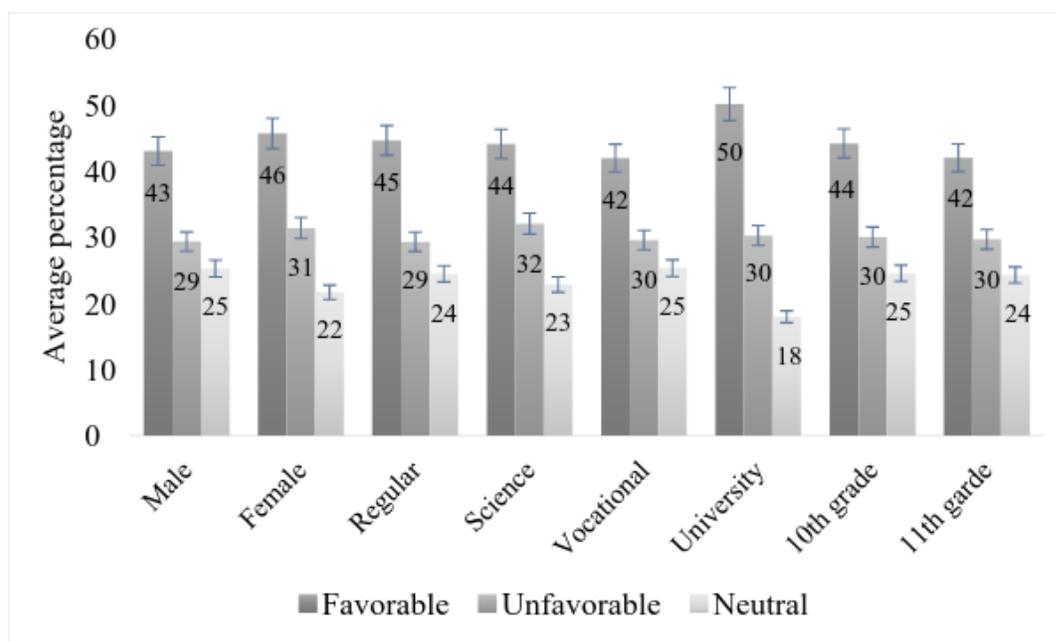

FIG. 1. Average percentages of favorable, unfavorable and neutral responses for various groups of Turkish students. Because of the blank responses, the sum of percentages for each group does not add up exactly to 100%. Error bars are calculated according to the standard errors for each group. The only statistically significant difference is between the university and vocational high school students' favorable responses.

Finally, the comparison of different high school grade levels in Fig. 1 shows that 44% and 42% of the 10[th] grade and 11[th] grade students, respectively, had an average expert-like response. However, the results of the t-test indicated that the differences between the averages of these two high school grade levels were neither significant for favorable (p=0.137) nor for unfavorable responses (p=0.777).

### B. Descriptive analysis for the survey items

In Fig. 1, various groups were compared based on average percentage favorable, unfavorable and neutral responses on the entire survey. Below, we analyze students' total scores for each item across different groups.

#### 1. Analysis based upon gender

Table AI in Appendix I shows the average normalized scores for all students separated by gender for each item. This Table shows more unfavorable average responses on eight items for male students (items 1, 3, 5, 11, 12, 16, 23, 30) and on nine items for female students (items 1, 3, 5, 6, 11, 12, 23, 30, 31). In fact, for seven common items, both male and female students have more unfavorable average responses. Question 12 is an item on which, for both genders, the most novice-like responses are observed (negative normalized score for females=-0.68, and for males=-0.52), and students of both genders were likely to agree with the statement that physics involves many equations each of which applies primarily to a specific situation.

On the other hand, the maximum normalized score for male students is on item 29 (+0.52), on which 65% of males noted that if their answer to a physics problem was not reasonable, they trace back their solution to see where they went wrong. The maximum normalized score for female students is on item 21 (+0.63), on which 73% of females claimed that after solving several physics problems in which the same principle is applied in different contexts, they should be able to apply the same principle in other situations.

Moreover, on item 6, in which students were asked if they can often tell if their work and/or answer is wrong even without external resources while solving physics problems, the average favorable response is larger for male students compared to female students. Similarly, on item 31, in which students were asked if they prefer to solve physics problems symbolically first before substituting values, the average favorable response is larger for male students compared to female students. On the other hand, on item 8, which asks about whether there is usually only one correct way to solve a given physics problem, the average favorable response is larger for female students compared to male students (see Table AI).

   *2. Analysis based upon School Type*

We also analyzed data based on school type to investigate whether average student responses to the survey items were different based upon the type of school. Table AII shows the normalized score of regular high school, science high school, and vocational high school and university students for each item. A closer look at Table AII in Appendix I indicates that for items 1, 3, 5, 11, 12, 23 and 30, the number of unfavorable responses exceeds that of favorable responses for all types of schools, including the university. What is interesting is that on items 1, 6, 11, 12, 30 and 31 the normalized scores of the university students are more unfavorable (negative) than high school students in all three types of high schools.

However, Table AII also shows that on some items (items 8, 16, 21, 22, 29), university students had a significantly more favorable response than high school students, indicating a higher level of expert-like attitudes and approaches to problem solving among university students as elicited by those survey items. For example, on item 16, while fewer high school students (regular=40%, science=47% and vocational=49%) claimed that they use their gut feeling to answer conceptual questions rather than invoking physics principles, a significantly higher percentage (56%) of the university students claimed to invoke physics principles in an expert-like approach to problem solving even on conceptual questions. In fact, Table AII shows that only the university students had a positive normalized score on item 16.

Table AII also shows that for the university students, the items with the most favorable (positive) normalized scores were items 21, 22 and 29 (these responses suggest that, on average, the university students strongly felt that they should be able to apply a physics principle in different contexts after they have solved several problems in which the principle is applied; if they obtained an answer to a physics problem that does not seem reasonable, they spend considerable time thinking about what may be wrong with the problem solution; and if they realize that their answer to a physics problem is not reasonable, they trace back their solution to see where they went wrong). Table AII also shows that for the university students, the items with the most unfavorable normalized scores (negative) were items 1, 11 and 12 (these responses suggest that, on average, if they were not sure about the right way to start a problem, they felt stuck without external help, they were more likely than high school students to also claim that equations are not things that one needs to understand in an intuitive sense, they routinely use equations to calculate numerical answers even if they are non-intuitive and physics involves many equations each of which applies primary to a specific situation). Moreover, for the high school students, the items with the most favorable normalized scores were items 7, 21, 28, 29 (followed by items 10, 22, 24, 25 and 26) and the items with the most unfavorable normalized scores (negative) were items 1, 3, 5, 11 and 12.

   *3. Analysis based upon Grade Level*

To analyze data in terms of grade level, there is no need to compare the university student responses with that of $10^{th}$ and $11^{th}$ grade students, since this comparison was done when comparing school types (see Table AII). Table AIII in Appendix I shows that there are some differences between the normalized scores of high school students in $10^{th}$ and $11^{th}$ grades, but the overall trend on each question is similar.

Table AIII shows that for items 1, 3, 5, 11, 12, 16, 23 and 30, the number of unfavorable responses exceeds that of favorable responses, so that the normalized item scores are negative for those items for both $10^{th}$ and $11^{th}$ grade students. For both grade levels, the most unfavorable normalized score is on item 12 in which 69% and 68% of $10^{th}$ and $11^{th}$ grade students, respectively, agreed with the novice-like statement that physics involves many equations each of which applies primarily to a specific situation . On the other hand, for $10^{th}$ and $11^{th}$ grades, the maximum normalized score (most expert-like response) was in response to item 29 and item 21, respectively. For example, in $10^{th}$ grade, 67% of students on item 29 claimed that if they realize that their answer to a physics problem is not reasonable, they trace back their solution to see where they went wrong and only 14% of them noted they would not do so. Similarly, on item 21, 62% of the 11[th] grade students claimed that after they solve several physics problems in which the same principle is applied in different contexts, they should be able to apply the same principle in other situations and only 15% of them noted that they would not be able to do so.

### C. Analysis on the basis of inferential statistics

To determine whether the differences between the groups on different survey items are statistically significant, the chi-square test for independence was used. As in the earlier discussions, participants were grouped by gender, school type and grade level and their responses to each of the 33 survey items were classified as favorable, unfavorable, or neutral. Table III shows the chi-square statistics for the items for which the frequency distributions of male and female students' favorable and unfavorable responses were statistically significantly different. In particular, although there is no statistically significant difference between the overall averages of male and female students, Table III shows that differences between the frequency distributions of male and female responses for 33% of the items were statistically significant.

Table III. Percentages of male and female students' favorable and unfavorable attitudes and approaches to problem solving and the corresponding chi-square statistics for items on which there is a statistically significant difference.

| Item | Male | | Female | | Chi square | | | |
|---|---|---|---|---|---|---|---|---|
| | favorable% | unfavorable% | favorable% | unfavorable% | $\chi^2$ | df | N | p |
| 1 | 24.7 | 54.1 | 11.2 | 74.7 | 21.3 | 2 | 514 | .000 |
| 3 | 23.9 | 58.0 | 22.5 | 67.5 | 6.5 | 2 | 512 | .039 |
| 6 | 45.1 | 20.9 | 33.7 | 40.2 | 21.3 | 2 | 513 | .000 |
| 8 | 38.9 | 36.8 | 61.2 | 21.8 | 22.0 | 2 | 512 | .000 |
| 12 | 13.5 | 65.2 | 10.5 | 78.4 | 10.3 | 2 | 513 | .006 |
| 16 | 25.7 | 49.7 | 48.8 | 31.2 | 28.3 | 2 | 516 | .000 |
| 19 | 48.8 | 22.4 | 60.9 | 19.5 | 7.3 | 2 | 499 | .026 |
| 21 | 60.5 | 15.4 | 72.7 | 9.7 | 7.3 | 2 | 497 | .026 |
| 23 | 24.0 | 50.8 | 27.1 | 58.2 | 7.3 | 2 | 503 | .025 |
| 31 | 45.3 | 24.0 | 31.8 | 41.2 | 16.7 | 2 | 503 | .000 |
| 33 | 48.8 | 27.3 | 39.9 | 22.7 | 9.7 | 2 | 489 | .008 |

Since data in Table III are on the basis of favorable and unfavorable responses, percentages for "neutral" responses are not displayed. In roughly half of the cases shown in Table III, the male responses are more favorable, and in the other half, female responses are more favorable. For example, for items 1, 3, 6, 12, 31 and 33, while the male responses are significantly more expert-like, for items 8, 16, 19, 21 and 23, the female responses demonstrate more expert-like attitude and approaches to problem solving. A closer look at the survey questions and Table III shows that male students' more expert-like responses were related to whether they feel stuck without external help if they are not sure about the right way to start a problem (item 1), whether they feel that being able to handle the mathematics is the most important part of the process in solving problems (item 3), whether they can tell whether the answer to a problem is correct or not without help (item 6), whether physics equations are applied primarily to a specific situation (item 12), whether they prefer to solve physics problems symbolically first (item 31) and whether two problems involving the same physics principle can be solved using similar methods even if the contexts are different (item 33). On the other hand, female students' expert-like responses were related to whether there is usually only one correct way to solve a given physics problem (item 8), whether they mostly use their gut feeling for answering conceptual questions unlike the solutions to quantitative problems for which they think of physics principles (item 16), whether different approaches should be used to answer a multiple-choice or a corresponding free response question (item 19), whether after applying the same physics principle to many situations, they should be able to apply it in other situations (item 21) and whether they give up solving a physics problem if they cannot solve it in 10 minutes (item 23). Moreover, Table III also shows that in terms of absolute differences between the average percent favorable responses, only two items (item 8 and item 16) had more than 20% difference between male and female students. On both these items, female students on average had more favorable responses than males.

Table IV shows the chi-square statistics for items for which the frequency distributions of regular, science, vocational and university student responses were statistically significantly different. As shown in Table IV, statistically significant differences between school types were evident for nine of the 33 items. For items 1 and 3, the percentage of unfavorable responses are statistically greater than favorable responses for all groups. In particular, for all school types, students responded that if they were not sure about the right way to start a problem, they would be stuck unless they got help (item 1) and students mostly perceived handling the mathematics as the most important part of the physics problem solving process (item 3). For items 8, 21, 22, and 29, the percentage of expert-like responses were greater for all groups when compared to novice-like responses. For example, Table IV shows that in response to item 22, many regular (52.9%), science (52.5%), and vocational (50.3%) high school students and university (89.7%) students claimed that they spend considerable time thinking about what may be wrong if they obtain an answer to a physics problem that does not seem reasonable. As noted earlier, on items 6 and 31, high school student responses were more expert-like than university students, but for item 16, the reverse is true.

Table IV. Percentages of regular, science, and vocational high school students' and university students' favorable (fav) and unfavorable (unfav) attitudes and approaches to problem solving and the corresponding chi-square statistics for significant items.

| | Regular | | Science | | Vocational | | Undergraduate | | Chi square | | | |
|---|---|---|---|---|---|---|---|---|---|---|---|---|
| Item | Fav. % | Unfav. % | Fav. % | Unfav. % | Fav. % | Unfav. % | Fav. % | Unfav. % | $\chi^2$ | df | N | p |
| 1 | 20.9 | 61.8 | 27.5 | 53.9 | 17.7 | 60.2 | 5.0 | 80.0 | 12.7 | 6 | 519 | .048 |
| 3 | 22.1 | 59.5 | 36.6 | 48.5 | 14.0 | 72.0 | 39.5 | 50.0 | 28.1 | 6 | 515 | .000 |
| 6 | 44.2 | 27.4 | 43.1 | 26.5 | 41.1 | 22.7 | 30.0 | 50.0 | 14.3 | 6 | 517 | .027 |
| 8 | 44.2 | 34.7 | 49.5 | 26.7 | 38.0 | 37.4 | 77.5 | 10.0 | 23.2 | 6 | 518 | .001 |
| 16 | 37.0 | 40.1 | 31.7 | 46.5 | 25.7 | 49.2 | 56.4 | 28.2 | 15.1 | 6 | 519 | .014 |
| 21 | 61.1 | 13.5 | 69.3 | 14.9 | 60.8 | 14.8 | 87.2 | 2.6 | 13.7 | 6 | 501 | .034 |
| 22 | 52.9 | 20.9 | 52.5 | 22.8 | 50.3 | 18.1 | 89.7 | 5.1 | 23.4 | 6 | 504 | .001 |
| 29 | 69.0 | 13.9 | 64.0 | 17.0 | 65.2 | 12.4 | 92.5 | 2.5 | 14.4 | 6 | 505 | .025 |
| 31 | 39.5 | 27.4 | 45.0 | 33.0 | 44.6 | 26.0 | 21.1 | 52.6 | 15.9 | 6 | 505 | .014 |

For grade-level, the statistically significant differences between the means of 10[th] and 11[th] grade students were observed only on item 9. On that item, 11[th] grade students were much more likely to provide an expert-like response (59%) than were the 10[th] grade students (47%), in terms of using a similar approach to solve all problems involving the same physics principle, even if the physical situations given in the problems are very different ($\chi^2$ =6.417, df=2, p=0 .040).

**D. Comparison of university students in Turkey with US university students in introductory algebra-based physics courses**

In the previous study, the AAPS survey was administered to 541 university students in the US in introductory physics courses [19]. Of this sample, 397 students were in first or second semester of introductory algebra-based physics courses. Although the students were not majoring in this same discipline, this group is most similar to the sample of 43 Turkish university students, who had taken algebra-based introductory mechanics and were enrolled in the second semester of an algebra-based physics course. The 397 students in the US were enrolled in different sections of first or second semester algebra-based introductory physics courses (all sections were pooled together since the average responses on the survey of students from different sections of these courses were not statistically significantly different) and they were mainly bioscience majors and pre-medical students [19]. The 43 Turkish university students, who were at the end of the second semester algebra-based electricity and magnetism course, were science education majors (in Turkey, this group of students is typically interested in becoming future middle school science teachers). We note that even though the sizes of the two groups and student majors are different, we compare the two groups to get some feel for students' attitudes and approaches to problem solving from different countries for somewhat similar groups, since both groups had students enrolled in university algebra-based physics courses[30-32].

The average normalized scores on the entire survey were +0.21 and +0.38, respectively, for Turkish and US students, which suggests that the US students' responses on average were more expert-like than those of Turkish students. On average, the Turkish university students' responses were unfavorable (negative normalized score on that item) on nine items (1, 3, 5, 6, 11, 12, 23, 30, 31), and the US students had unfavorable responses on only five items (11, 12, 20, 27, 30). Only in responding to items 11, 12 and 30, both groups on average displayed non expert-like attitudes and approaches to problem solving. For example, on those survey questions, students from both countries were more likely to claim that they routinely use equations to calculate numerical answers even if they are non-intuitive, that physics involves many equations each of which applies primarily to a specific situation, and that it is much more difficult to solve a physics problem with symbols than solving an identical problem with a numerical answer.

For comparing both groups' total scores, an independent sample t-test was carried out which shows that US students' responses were more expert-like on average ($p<0.05$) than Turkish students' responses. There is variability in the scores of both groups across various items so a Chi square analysis was performed. Statistically significant differences were found on items shown in Table V.

Table V. Percentages of Turkish and US University (algebra-based courses) students' favorable and unfavorable attitudes and approaches to problem solving and corresponding Chi-square statistics for items on which there is statistically significant difference. Note that the total number of students N is different between questions due to a few blank responses for some questions.

| Item | Turkish | | US | | Chi square | | | |
|---|---|---|---|---|---|---|---|---|
| | favorable% | unfavorable% | favorable% | unfavorable% | $\chi^2$ | df | N | p |
| 1 | 5.0 | 80.0 | 51.8 | 33.3 | 37.8 | 2 | 436 | .000 |
| 4 | 36.8 | 36.8 | 60.1 | 22.5 | 7.8 | 2 | 434 | .021 |
| 5 | 20.5 | 61.5 | 47.6 | 32.7 | 14.2 | 2 | 436 | .001 |
| 6 | 30.0 | 50.0 | 49.5 | 28.5 | 8.4 | 2 | 436 | .015 |
| 9 | 72.5 | 12.5 | 42.4 | 18.7 | 13.7 | 2 | 436 | .001 |
| 11 | 22.5 | 75.0 | 30.8 | 36.9 | 24.8 | 2 | 436 | .000 |
| 12 | 7.5 | 80.0 | 35.6 | 41.4 | 22.5 | 2 | 436 | .000 |
| 15 | 56.4 | 25.6 | 80.1 | 9.6 | 12.7 | 2 | 435 | .002 |
| 17 | 43.6 | 25.6 | 72.8 | 15.4 | 16.0 | 2 | 436 | .000 |
| 18 | 43.6 | 30.8 | 78.3 | 11.8 | 23.0 | 2 | 436 | .000 |
| 19 | 67.5 | 17.5 | 85.9 | 8.1 | 9.3 | 2 | 437 | .009 |
| 20 | 56.4 | 23.1 | 23.9 | 45.1 | 19.2 | 2 | 436 | .000 |
| 22 | 89.7 | 5.1 | 64.2 | 13.6 | 10.5 | 2 | 436 | .005 |
| 23 | 27.5 | 47.5 | 58.9 | 23.2 | 15.9 | 2 | 437 | .000 |
| 30 | 20.0 | 57.5 | 38.0 | 38.5 | 6.5 | 2 | 437 | .039 |
| 31 | 21.1 | 52.6 | 43.8 | 38.3 | 7.4 | 2 | 435 | .024 |
| 32 | 61.5 | 25.6 | 81.8 | 12.4 | 9.1 | 2 | 434 | .011 |
| 33 | 51.3 | 12.8 | 69.2 | 19.0 | 17.1 | 2 | 429 | .000 |

Table V shows that on more than half of the items (18 items), the responses of the US and Turkish university students in algebra-based introductory physics courses were statistically significantly different. In terms of expert-like attitudes and approaches to problem solving, these statistically significant differences are in favor of the Turkish students only on items 9, 20 and 22. For example, the Turkish students' responses suggest that they are more likely to use a similar approach to solving all problems involving a physics principle even if the physical situations given in the problems are very different; after solving homework problems, they are more likely to take the time to reflect and learn from the solution; and if they obtain an answer to a physics problem that does not seem reasonable, they are more likely to spend considerable time thinking about what may be wrong with the problem solution. On the other hand, the US students claimed to have significantly more expert like attitudes and approaches to problem solving on the other items. For example, on item 1, they were less likely to feel stuck unless they got help if they were not sure about the right way to start a problem; on item 4, they were more likely to claim that in solving physics problems, they always identified the physics principles involved in the problem first before looking for corresponding equations; on item 6, they were more likely to claim that they can often tell when their work and/or answer to a physics problem is wrong even without external help; on item 15, they often find it useful to first draw a picture of the situations described in the physics problems; on item 19, they were equally likely to do scratch work when answering a multiple-choice question or a corresponding free-response question; and on items 32 and 33, they were more likely to solve different problems involving the same principle using similar methods even if the contexts were very different. We note that the researchers compared typical midterm and final exams in algebra-based university physics courses in the US and Turkey. While the questions are on similar topics, the questions in Turkish exam appear to be somewhat more difficult than the exams administered to the US students. This difference could be one possible reason why more Turkish students agree with the statement (on item 1) that they feel stuck unless they get external help.

### E. Factor analysis

Ref. [5] lists the pros and cons of confirmatory factor analysis and exploratory factor analysis. An exploratory factor analysis was used to find the relationships and patterns among AAPS survey items [34]. The data from 528 participants, which easily meets the recommended sample size of at least 300 participants [34], was used to group the items. For suitability of our data for factor analysis, we used the Kaiser-Meyer-Olkin (KMO) measure of sampling adequacy and Bartlett's test to confirm that our example has patterned relationships. The KMO value was found to be 0.812 and Bartlett's test was found to be statistically significant ($\chi^2$=2122.641, df=528, p=.000). Since all requirements are met, distinct and reliable factors can be expected from our sample.

All 33 items of the AAPS survey were used for principal components analysis while performing varimax axis rotation. A total of 10 factors (composed of groups of six to two items) were obtained that explained 51.86% of the total variance. The variance explained by the scale indicates that AAPS survey measured the students' attitudes and approaches to physics problems adequately. All of the items on the AAPS survey had factor loadings that were greater than the lower limit of 0.30, ranging between 0.329 and 0.758. Thus, all AAPS survey items are likely to make a meaningful contribution to the survey [35].

Table AIV in Appendix II presents the findings of the exploratory factor analysis in detail; in which each factor has a description that summarizes the common link between questions in that factor. Two researchers came up with the descriptions separately, and then all three of researchers discussed the descriptions and jointly agreed on the descriptions in Appendix II after discussions. Some of the factors focus on attitudes and approaches to problem solving in specific cases (e.g., drawing diagrams and doing scratch work) while others focus on boarder issues (e.g., metacognition). For example, factor 1 contains the largest number of items and relates to metacognition in physics problem solving. The items in factor 4 are very closely connected, e.g. three of the items associated with this factor relate to utility of drawing diagrams/pictures and scratch work. While most factors were rather straightforward to describe and the researchers came up with very similar descriptions of the factors (see Appendix II), factor 2 required discussion between the researchers before finalizing on a description of the factor. In particular, factor 2 features questions that investigate student views about whether there is only one way to solve a physics

problem, whether one should use gut feeling for conceptual questions, and whether one should use a symbolic solution first for numerical problems or not. After deliberation, the researchers agreed that the common link appears to be novice-like approaches to problem solving.

## IV. DISCUSSION AND SUMMARY

We describe the validation, administration and analysis of data for the Turkish version of the AAPS survey for high school and university students in Turkey. The analysis of data was conducted by grade level, school type, and gender. The comparison of the responses of the Turkish students enrolled in an algebra-based university introductory physics course with responses of the American students enrolled in equivalent courses in the US is also made.

We find that the Turkish version of the AAPS survey which was validated with experts and students has a robust internal consistency for a large sample of Turkish high school and university students, regardless of grade level, school type, or gender. In other words, the survey highlights of the main findings are summarized in Table VI.

Table VI. Highlights of the main findings are summarized

| |
|---|
| • No statistically significant differences were observed on average normalized gains between gender, grade level and type of high school. (All 33 items considered together) |
| • The only statistically significant difference was that the university students answered more favorably than vocational high school students when the response categories are considered favorable, neutral and unfavorable. |
| • On an item by item basis, statistically significant differences between gender groups were observed on 11 items (1, 3, 6, 8, 12, 16, 19, 21, 23, 31, 33 items). On items 1, 3, 6, 23, 31 and 33, male students were more expert-like than female students. On the other hand, on items 8, 12, 16, 19 and 2, female students were more expert-like than male students. |
| • Significant differences between school types were observed on nine items (1, 3, 6, 8, 16, 21, 22, 29, 31). On items 1, 3, 6 and 31, high school students were more expert-like than university students, but on items 8, 16, 21, 22 and 29, the reverse condition is valid. |
| • No statistically significant differences between the averages of 10th and 11th grade students were observed either overall on the entire survey or on individual items, with the exception of item 9 (on which 11$^{th}$ grade students' responses were more expert-like). |
| • The principal component analysis for the 33 items of AAPS yielded a total of 10 factors that explained 51.86% of the total variance. Metacognition in physics problem solving was the description of the factor with the largest number of items. |
| • On more than half of the items (18 items), the responses of the US and Turkish university students in algebra-based introductory physics courses were statistically significantly different, with Turkish students performing more expert-like on three items and US students performing more expert-like on the other fifteen items. |

When comparing the average scores for all 33 questions together, the Turkish students from different comparison groups (based upon gender, school type or grade level) show similar trends of favorability in their responses. However, one statistically significant result is that the Turkish university students' responses are more expert-like than vocational high school students' responses, which may partly be explained by the relatively lower priority that a vocational school curriculum places on learning physics compared to that of other high schools, and the fact that vocational high school students typically perform worst among all students in the national high school entrance tests (including on the science part). No other comparisons for the overall average survey score between grade levels or between school types are statistically significant. This is consistent with the expectation that high school students and university students in their first year are relatively novice-like in their attitudes and approaches to problem solving, so their responses on the AAPS survey overall are likely to be similar.

For individual items on the survey, interesting trends emerge in different comparison groups. For example, between different grade levels, certain items (items 8, 16, 21, 22, 29) were answered with a significantly more favorable response by university students than by high school students, indicating a higher level of expert-like attitudes and approaches to problem solving among university students as elicited by those survey items.

On the other hand, university students demonstrated a more novice-like approach on some items (items 1, 6, 11, 12, 30 and 31) compared to high school students, suggesting that there may be some differences between the two levels of schooling, or other factors that may adversely influence attitudes and approaches of university students on problem solving pertaining to those specific topics. Since our study was not designed to investigate the reason for these interesting differences, future studies will investigate the possible reasons for these differences. However, we note that of these items, items 11 and 12 are in the $5^{th}$ factor, item 31 is in the $6^{th}$ factor, items 1 and 30 are in the $7^{th}$ factor, and item 6 is in the $9^{th}$ factor of the principal component analysis presented in the Appendix II. A closer look at the description of these factors in the Appendix II suggests that the Turkish university students demonstrated more novice-like attitudes and approaches than high school students on items related to the role of equations/formulas, solving difficult problems, and knowing when their solution is wrong. One possible reason for the worse responses of the university students compared to high school students is that students in grade 10 and 11 generally solve more conceptual problems than university students, and the types of problems the high school students generally solve are also lower on Bloom's taxonomy than those that the university students solve. Since the university students start to deal with more difficult, more quantitative problems, it is possible that they are more likely to think that physics mainly consists of equations and formulas and they may have reduced confidence in being able to solve difficult problems and knowing when their solution is wrong.

Regarding the effect of gender on student responses, although there were statistically significant differences between males and females on several items, in some cases female student responses were more favorable and in other cases, male student responses were more favorable. Moreover, the differences between average male and female student favorable responses were larger than 20% only on two items with more favorable responses from female students on both of those problems.

An exploratory factor analysis using the principal component method suggests that all 33 questions provide a meaningful contribution to the AAPS survey. The most prevalent factor, in terms of the number of items, appears to be related to metacognition in physics problem solving.

The university-level students in Turkey majoring in science education enrolled in the second semester of an algebra-based introductory physics course were compared to the university-level algebra-based introductory physics students in the United States from the original Mason and Singh study [19]. The average normalized score for the entire survey is almost double for the US students compared to the Turkish students and a significant difference exists between American and Turkish students on many questions, more often in favor of American students. This study does not investigate why US university students in the algebra-based introductory physics courses in general performed better than the Turkish university students on the survey. Apart from the differences in what the students were majoring in, the large average differences on survey responses may at least partly be due to the social and cultural differences between the two countries and the differences in their educational systems and assessment tools and methods. A comparison of the exams administered in these courses in the US and Turkey suggests that while the exams cover similar concepts, the Turkish exams are somewhat more difficult. This difference in the difficulty of the exams can also impact students' attitudes and approaches, especially if there is inadequate guidance and support to help students learn physics and develop useful skills to perform well. Another potential reason for the Turkish students doing significantly worse than the American students on most survey questions on which there were significant differences may be that the Turkish students answered the survey questions more honestly than the American students. This difference may partly be because Turkish students were not given any bonus course credit for answering the survey questions but were told that this was a research survey to improve education, whereas American students were given some bonus course credit. Even though US students were told that the bonus course credit was not dependent on their actual responses to the survey items, some students may still have answered the questions in a more expert-like manner because they may have worried that the instructor may correlate their survey responses with their actual performance on problem solving. Individual interviews conducted with a subset of US students, [18] in which they were asked to solve physics problems along with answering the survey questions, also suggests that their survey responses were generally more favorable than their actual attitudes and approaches while solving problems. These types of issues about the reasons for the differences between the average responses of students in the US and Turkey will be investigated in the future studies.

## ACKNOWLEDGEMENTS

We thank all of the faculty, postdocs, and students at various levels who helped us during the validation of the Turkish version of the survey. We also thank Fred Reif and Chris Schunn for extremely helpful discussions related to data analysis and statistics.

## APPENDIX I

Table AI. Normalized scores on each item for male and female students.

| # | 1 | 2 | 3 | 4 | 5 | 6 | 7 | 8 | 9 | 10 | 11 |
|---|---|---|---|---|---|---|---|---|---|---|---|
| Female | -0.64 | +0.16 | -0.45 | +0.23 | -0.45 | -0.07 | +0.54 | +0.39 | +0.39 | +0.39 | -0.46 |
| Male | -0.29 | +0.19 | -0.34 | +0.28 | -0.36 | +0.24 | +0.45 | +0.02 | +0.33 | +0.41 | -0.41 |
| # | 12 | 13 | 14 | 15 | 16 | 17 | 18 | 19 | 20 | 21 | 22 |
| Female | -0.68 | +0.35 | +0.24 | +0.25 | +0.18 | +0.17 | +0.24 | +0.41 | +0.28 | +0.63 | +0.42 |
| Male | -0.52 | +0.23 | +0.23 | +0.27 | -0.24 | +0.26 | +0.19 | +0.26 | +0.29 | +0.45 | +0.32 |
| # | 23 | 24 | 25 | 26 | 27 | 28 | 29 | 30 | 31 | 32 | 33 |
| Female | -0.31 | +0.43 | +0.39 | +0.34 | +0.25 | +0.51 | +0.61 | -0.25 | -0.09 | +0.25 | +0.17 |
| Male | -0.27 | +0.31 | +0.41 | +0.35 | +0.24 | +0.44 | +0.52 | -0.24 | +0.21 | +0.23 | +0.22 |

Table AII. Normalized scores on each item for different school types

| # | 1 | 2 | 3 | 4 | 5 | 6 | 7 | 8 | 9 | 10 | 11 |
|---|---|---|---|---|---|---|---|---|---|---|---|
| Regular | -0.41 | +0.30 | -0.37 | +0.31 | -0.46 | +0.17 | +0.47 | +0.10 | +0.26 | +0.46 | -0.42 |
| Science | -0.27 | +0.02 | -0.12 | +0.17 | -0.31 | +0.17 | +0.50 | +0.23 | +0.28 | +0.32 | -0.42 |
| Vocational | -0.43 | +0.15 | -0.58 | +0.33 | -0.37 | +0.18 | +0.44 | +0.01 | +0.42 | +0.39 | -0.40 |
| University | -0.75 | +0.10 | -0.11 | 0.00 | -0.41 | -0.20 | +0.68 | +0.68 | +0.60 | +0.39 | -0.53 |
| # | 12 | 13 | 14 | 15 | 16 | 17 | 18 | 19 | 20 | 21 | 22 |
| Regular | -0.57 | +0.33 | +0.25 | +0.31 | -0.03 | +0.209 | +0.25 | +0.39 | +0.27 | +0.48 | +0.32 |
| Science | -0.57 | +0.28 | +0.09 | +0.23 | -0.15 | +0.07 | +0.14 | +0.18 | +0.17 | +0.55 | +0.30 |
| Vocational | -0.54 | +0.19 | +0.20 | +0.24 | -0.24 | +0.38 | +0.23 | +0.29 | +0.39 | +0.46 | +0.32 |
| University | -0.73 | +0.50 | +0.26 | +0.30 | +0.28 | +0.18 | +0.13 | +0.50 | +0.33 | +0.85 | +0.85 |
| # | 23 | 24 | 25 | 26 | 27 | 28 | 29 | 30 | 31 | 32 | 33 |
| Regular | -0.28 | +0.33 | +0.37 | +0.44 | +0.28 | +0.51 | +0.55 | -0.16 | +0.12 | +0.21 | +0.20 |
| Science | -0.18 | +0.43 | +0.41 | +0.26 | +0.27 | +0.45 | +0.47 | -0.35 | +0.12 | +0.25 | +0.08 |
| Vocational | -0.38 | +0.34 | +0.40 | +0.30 | +0.23 | +0.40 | +0.53 | -0.26 | +0.19 | +0.21 | +0.23 |
| University | -0.20 | +0.30 | +0.63 | +0.41 | +0.13 | +0.53 | +0.90 | -0.38 | -0.31 | +0.35 | +0.38 |

Table AIII. Normalized scores on each item for different grade levels

| # | 1 | 2 | 3 | 4 | 5 | 6 | 7 | 8 | 9 | 10 | 11 |
|---|---|---|---|---|---|---|---|---|---|---|---|
| 10th grade | -0.35 | +0.17 | -0.46 | +0.29 | -0.39 | +0.17 | +0.50 | +0.06 | +0.26 | +0.45 | -0.41 |
| 11th grade | -0.45 | +0.24 | -0.31 | +0.27 | -0.43 | +0.18 | +0.39 | +0.12 | +0.43 | +0.31 | -0.38 |
| # | 12 | 13 | 14 | 15 | 16 | 17 | 18 | 19 | 20 | 21 | 22 |
| 10th grade | -0.54 | +0.22 | +0.28 | +0.31 | -0.14 | +0.30 | +0.24 | +0.30 | +0.25 | +0.47 | +0.36 |
| 11th grade | -0.56 | +0.31 | +0.17 | +0.21 | -0.12 | +0.17 | +0.17 | +0.32 | +0.32 | +0.51 | +0.25 |
| # | 23 | 24 | 25 | 26 | 27 | 28 | 29 | 30 | 31 | 32 | 33 |
| 10th grade | -0.27 | +0.32 | +0.38 | +0.40 | +0.28 | +0.48 | +0.52 | -0.19 | +0.14 | +0.16 | +0.20 |
| 11th grade | -0.28 | +0.39 | +0.37 | +0.26 | +0.25 | +0.40 | +0.48 | -0.27 | +0.16 | +0.31 | +0.15 |

**APPENDIX II**

Table AIV: Results of the Principal Component Analysis along with a description of each factor.

| Factor (Variance explained) | Item | Loading | Description |
|---|---|---|---|
| Factor1 (6.856) | 22 | .633 | Metacognition in physics problem solving |
| | 7 | .554 | |
| | 21 | 497 | |
| | 29 | .463 | |
| | 10 | 376 | |
| | 13 | .329 | |
| Factor2 (6.668) | 4 | .698 | Connections to physics concepts and the real world |
| | 2 | .614 | |
| | 3 | .534 | |
| | 14 | .510 | |
| | 20 | .390 | |
| Factor3 (5.701) | 26 | .762 | Enjoyment and utility of solving challenging physics problems |
| | 27 | .622 | |
| | 28 | .509 | |
| Factor4 (5.555) | 18 | .758 | Utility of drawing pictures and/or diagrams or scratch work in physics problem solving |
| | 17 | .690 | |
| | 19 | .601 | |
| Factor5 (5.028) | 11 | .598 | Role of equations/formulas in physics problem solving |
| | 5 | .580 | |
| | 12 | .579 | |
| Factor6 (4.849) | 8 | .720 | Novice-like approaches to physics problem solving |
| | 31 | .517 | |
| | 16 | .441 | |
| Factor7 (4.470) | 30 1 | .714 .527 | Views towards - difficult problems |
| Factor8 (4.376) | 25 24 15 | .552 .479 .365 | Sense-making (Effective general strategies for solving and learning from problems) |
| Factor9 (4.273) | 23 6 | .666 -.605 | Problem solving confidence (Knowing when the solution is wrong and not giving up) |
| Factor10 (4.083) | 32 33 9 | .707 .603 .420 | Solving different problems using the same principle |